\documentstyle[12pt,epsf]{article}
\catcode`\@=11
\long\def\@makefntext#1{
\protect\noindent \hbox to 3.2pt {\hskip-.9pt
$^{{\ninerm\@thefnmark}}$\hfil}#1\hfill}                

\def\@makefnmark{\hbox to 0pt{$^{\@thefnmark}$\hss}}  

\def\ps@myheadings{\let\@mkboth\@gobbletwo
\def\@oddhead{\hbox{}
\rightmark\hfil\ninerm\thepage}
\def\@oddfoot{}\def\@evenhead{\ninerm\thepage\hfil
\leftmark\hbox{}}\def\@evenfoot{}
\def\sectionmark##1{}\def\subsectionmark##1{}}

\renewcommand{\thefootnote}{\fnsymbol{footnote}}

\newcounter{sectionc}\newcounter{subsectionc}\newcounter{subsubsectionc}
\renewcommand{\section}[1] {\vspace*{0.6cm}\addtocounter{sectionc}{1}
\setcounter{subsectionc}{0}\setcounter{subsubsectionc}{0}\noindent
        {\normalsize\bf\thesectionc. #1}\par\vspace*{0.4cm}}
\renewcommand{\subsection}[1] {\vspace*{0.6cm}\addtocounter{subsectionc}{1}
        \setcounter{subsubsectionc}{0}\noindent
        {\normalsize\it\thesectionc.\thesubsectionc. #1}\par\vspace*{0.4cm}}
\renewcommand{\subsubsection}[1]
{\vspace*{0.6cm}\addtocounter{subsubsectionc}{1}
        \noindent {\normalsize\rm\thesectionc.\thesubsectionc.\thesubsubsectionc.
        #1}\par\vspace*{0.4cm}}

\newcounter{appendixc}
\newcounter{subappendixc}[appendixc]
\newcounter{subsubappendixc}[subappendixc]

\renewcommand{\appendix}[1] {\vspace*{0.6cm}
        \refstepcounter{appendixc}
        \setcounter{figure}{0}
        \setcounter{table}{0}
        \setcounter{equation}{0}
        \renewcommand{\thefigure}{\Alph{appendixc}.\arabic{figure}}
        \renewcommand{\thetable}{\Alph{appendixc}.\arabic{table}}
        \renewcommand{\theappendixc}{\Alph{appendixc}}
        \renewcommand{\theequation}{\Alph{appendixc}.\arabic{equation}}
        \noindent{\bf Appendix \theappendixc #1}\par\vspace*{0.4cm}}

\def\abstracts#1{{
        \centering{\begin{minipage}{12.2truecm}\footnotesize\baselineskip=12pt\noindent
        \centerline{\footnotesize ABSTRACT}\vspace*{0.3cm}
        \parindent=0pt #1
        \end{minipage}}\par}}

\renewenvironment{thebibliography}[1]
        {\begin{list}{\arabic{enumi}.}
        {\usecounter{enumi}\setlength{\parsep}{0pt}
\setlength{\leftmargin 1.25cm}{\rightmargin 0pt}
         \setlength{\itemsep}{0pt} \settowidth
        {\labelwidth}{#1.}\sloppy}}{\end{list}}

\newcounter{itemlistc}
\newcounter{romanlistc}
\newcounter{alphlistc}
\newcounter{arabiclistc}

\newcommand{\fcaption}[1]{
        \refstepcounter{figure}
        \setbox\@tempboxa = \hbox{\footnotesize Fig.~\thefigure. #1}
        \ifdim \wd\@tempboxa > 6in
           {\begin{center}
        \parbox{6in}{\footnotesize\baselineskip=12pt Fig.~\thefigure. #1}
            \end{center}}
        \else
             {\begin{center}
             {\footnotesize Fig.~\thefigure. #1}
              \end{center}}
        \fi}

\newcommand{\tcaption}[1]{
        \refstepcounter{table}
        \setbox\@tempboxa = \hbox{\footnotesize Table~\thetable. #1}
        \ifdim \wd\@tempboxa > 6in
           {\begin{center}
        \parbox{6in}{\footnotesize\baselineskip=12pt Table~\thetable. #1}
            \end{center}}
        \else
             {\begin{center}
             {\footnotesize Table~\thetable. #1}
              \end{center}}
        \fi}

\def\@citex[#1]#2{\if@filesw\immediate\write\@auxout
        {\string\citation{#2}}\fi
\def\@citea{}\@cite{\@for\@citeb:=#2\do
        {\@citea\def\@citea{,}\@ifundefined
        {b@\@citeb}{{\bf ?}\@warning
        {Citation `\@citeb' on page \thepage \space undefined}}
        {\csname b@\@citeb\endcsname}}}{#1}}

\newif\if@cghi
\def\cite{\@cghitrue\@ifnextchar [{\@tempswatrue
        \@citex}{\@tempswafalse\@citex[]}}
\def\citelow{\@cghifalse\@ifnextchar [{\@tempswatrue
        \@citex}{\@tempswafalse\@citex[]}}
\def\@cite#1#2{{$\null^{#1}$\if@tempswa\typeout
        {IJCGA warning: optional citation argument
        ignored: `#2'} \fi}}

 1
 1
 1

\font\tenrm=cmr10
\font\tenit=cmti10

\font\ninerm=cmr9

\begin{document}
June 1995\hspace*{\fill}HLRZ 95-30~~~~~~~\\
\hspace*{\fill}WUB 95-20~~~~~~\vspace{1cm}\\

\centerline{\Large\bf Quark-Antiquark Forces From SU(2) And SU(3)}

\vspace{.3cm}

\centerline{\Large\bf  Gauge Theories On Large Lattices}

\vspace{1.5cm}

\centerline{\large\rm Gunnar S.\ Bali\footnote{Invited talk given
at ``International Workshop on Color Confinement and Hadrons,
CONFINEMENT 95'', March 22--24, 1995, RCNP Osaka, Japan.},
Klaus Schilling, Armin
Wachter}

\vspace{.8cm}

\baselineskip=13pt
\centerline{\tenit HLRZ c/o KFA, 52425 J\"ulich, Germany and DESY,
Hamburg, Germany}
\centerline{\tenrm and}
\centerline{\tenit Fachbereich Physik, Bergische Universit\"at,
42097 Wuppertal, Germany}

\vspace{4cm}
\abstracts{We present results on the spin-independent
quark-antiquark potential in SU(3) gauge theory from a simulation on
a $48^3 \times 64$ lattice at $\beta = 6.8$, corresponding to
a volume of $(1.7~\mbox{fm})^3$. Moreover, a comprehensive analysis 
of spin- and velocity-dependent potentials is carried out
for SU(2) gauge theory, with emphasis on the short range 
structure, on lattices with resolutions ranging from
$.02$~fm to $.04$~fm.}

\newpage
\normalsize\baselineskip=15pt
\setcounter{footnote}{0}
\renewcommand{\thefootnote}{\alph{footnote}}

\section{Introduction}
Quarkonia spectroscopy provides a wealth of information and thus
constitutes an important observational window to the phenomenology of
strong interactions. Within their non- (or semi-) relativistic
setting, potential models have been proven to describe the empirical
charmonium and bottomonium spectra remarkably well.
The resulting phenomenological potentials 
can  be compared to lattice predictions from the
QCD Lagrangian.

Lattice QCD techniques offer on the other hand a direct access
to the hadron spectrum, without recourse to the potential picture.
In the heavy quark sector (with quark mass $M$), however,
the full fledged
lattice calculation would require (at present) prohibitively fine lattice
resolutions $a<M^{-1}$. It is therefore more practical to expand
the QCD Lagrangian in powers of $M^{-1}$ into an effective
non-relativistic QCD action (NRQCD) which can be evaluated
subsequently
by use of lattice methods.\cite{NRQCD}

The traditional semi-relativistic potential model approach to
spectrum calculations assumes the validity of the instantaneous
approximation.  This approximation can be established from first
principles lattice computations by comparing spectra and wave functions
as obtained from NRQCD to results from lattice potentials.

Pioneering attempts to compute corrections to the static potential on
the lattice have been launched in the mid eighties.\cite{early,michael} In the
meantime, tremendous progress has been achieved, both in computational
power and methods. In view of the general interest in the potential
formulation of quark binding problems, it appears to be timely for
a new lattice determination of spin-dependent (sd) and
velocity-dependent (vd) potentials. 

As a first step into this direction, we present recent results on the
static quark antiquark potential from SU(3) gauge theory, as well as
results on sd and vd potentials from SU(2) gauge theory.  The
limitation to two colours will not yet allow to proceed to spectrum
calculations but will hopefully disclose the key features of
confinement at work.

\section{The Central Quark-Antiquark Potential}
In the limit of infinitely heavy colour sources
($M\rightarrow\infty$), the Born-Oppenheimer approximation can be applied
and ---  after integrating out the gauge degrees of freedom ---
the underlying QCD binding problem becomes nonrelativistic.

Theoretically, one expects the leading order (in $M^{-1}$) potential
to be  dominated by one-gluon exchange at short distances,
$V_0(r)=-C_F\frac{\alpha(r)}{r}$,  with a running coupling
$\alpha(r)$ which depends logarithmically on $r$.
On the other hand, for large separations $r$,
one would expect a 
 bosonic string model to hold which predicts the asymptotic
behaviour \cite{}
$V_0(r)=\kappa r-\frac{\pi}{12}\frac{1}{r}+\cdots$
(with string tension $\kappa$).
{\em A priori} it is not clear when {\it large} or {\it small}
$r$ behaviour would set in and which form the potential takes in the
region of intermediate $r$.

\begin{figure}
\centerline{\mbox{\epsfxsize=12cm\epsfbox{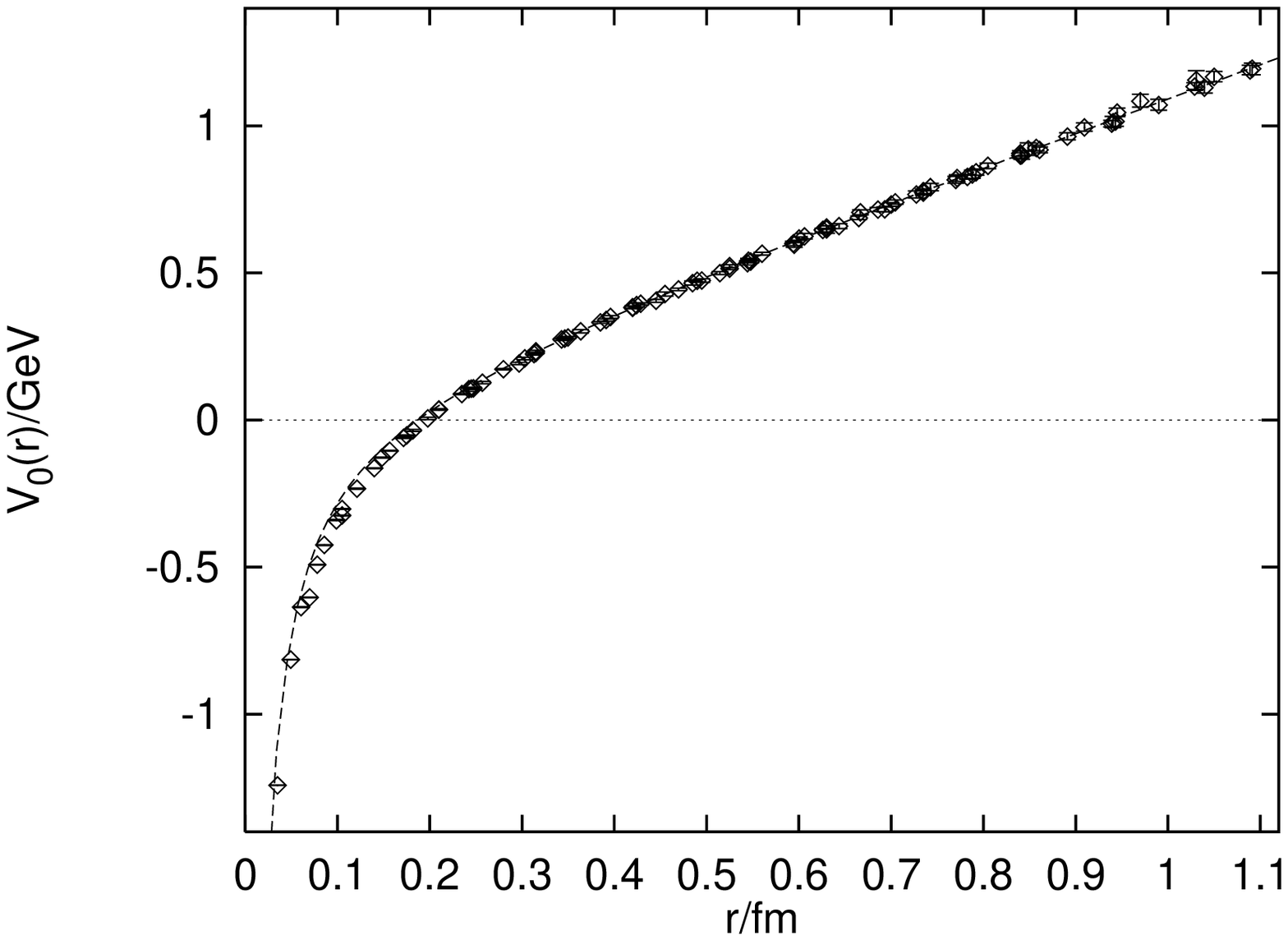}}}
\fcaption{The static SU(3) $Q\bar{Q}$ potential, obtained on a
$48^3\times 64$ lattice at $\beta=6.8$, corresponding to a lattice
spacing $a\approx .035$~fm.}
\label{fig:pot}
\end{figure}

Phenomenologically, the Cornell potential
$V_{phen}(r)=\kappa r-\frac{e}{r}$ has been proven to describe the
spin averaged charmonium and bottomonium spectra
successfully\cite{cornell} with one set of parameters, $\kappa$ and $e$,
in accord with the flavour independence of strong interactions.  It is
well known, though, that the binding energies of low lying states are
not overly sensitive to the shape of the potential 
 at distances smaller than
0.2~fm or larger than 1~fm.

Therefore, a lattice determination of the QCD
potential in the intermediate $r$ region is of particular interest. In
Fig.~\ref{fig:pot} we show high statistics results on the central
potential from SU(3) gauge theory on a $48^3\times 64$ lattice at
$\beta=6.8$, which corresponds to a lattice resolution $a\approx
0.035$~fm. The data for $r\geq 0.3$~fm has been fitted to a Cornell
type parametrization (solid curve). At small $r$ we observe deviations
{}from this parametrization. This is in accord with running coupling
effects.\cite{oldie} Note, that in the lattice computation vacuum
polarization due to sea quarks
has been neglected and thus no string breaking occurs.

Before one can proceed  to predict spin-averaged quarkonia spectra from
such lattice potentials, one has  to investigate possible
corrections to the infinite mass limit
at realistic charm and bottom quark masses. In solving the
Schr\"odinger equation with our potential, we find
the average speed of the quarks (in units of the speed of light)
to be $\langle v^2\rangle\approx 0.25$
and $\langle v^2\rangle\approx 0.09$ for the charmonium and
bottomonium ground states, respectively. This leads
us to expect that, at least in the case 
of charmonium, the phenomenological potential, $V_{phen}(r)$,
(which has been
tuned to reproduce the empirical  spectrum) might deviate  by substantial
${\cal O}(v^2)$ corrections
{}from the static potential, $V_0(r)$.
Needless to say, that fine and hyperfine splittings
have their origin in such $v$ or, equivalently, $1/M$ corrections.

\section{$1/M$-Corrections}
Starting from a Foldy-Wouthuysen transformation of the Euclidean
quark propagator in an external gauge field, the asymptotic
($t\rightarrow\infty$) expression
$W(r,t)\propto\exp(-V_0(r)t)$ can be derived for the static
potential, $V_0(r)$, where $W(r,t)$ denotes the expectation value
of the familiar Wilson loop
with spatial extent $r$ and temporal extent $t$.
Perturbing the propagator in terms of the
inverse quark masses $M_1^{-1}$ and $M_2^{-1}$ around its static
solution, one arrives at the semi-relativistic Hamiltonian (in the CM
system, ${\mathbf p_1}=-{\mathbf p_2}={\mathbf p}$),
\begin{equation}
H=\frac{{\mathbf p}^2}{2}\left(\frac{1}{M_1}+\frac{1}{M_2}\right)
-\frac{\left({\mathbf p}^2\right)^2}{8}
\left(\frac{1}{M_1^3}+\frac{1}{M_2^3}\right)
+V(r,{\mathbf L},{\mathbf S_1},{\mathbf S_2},
{\mathbf p})\quad,
\end{equation}
where the potential consists of a central part, sd and vd
corrections,
\begin{equation}
V(r,{\mathbf L},{\mathbf S_1},{\mathbf S_2},
{\mathbf p})=V_0(r)+V_{sd}(r,{\mathbf L},{\mathbf S_1},{\mathbf S_2})+
V_{vd}(r,{\mathbf p})\quad.
\end{equation}
The sd contributions have been derived by Eichten, Feinberg and
Gromes\cite{EFG} while
the vd terms have been elaborated by Barchielli {\em et al.}:\cite{VDEP}
\begin{eqnarray}
V_{sd}(r,{\mathbf L},{\mathbf S_1},{\mathbf S_2})
&=&\frac{{\mathbf l_1}{\mathbf s_1} + {\mathbf l_2}{\mathbf s_2}}
{2r}\left(V_0'(r)+2V_1'(r)\right)
+\frac{{\mathbf l_1}{\mathbf s_2} + {\mathbf l_2}{\mathbf
s_1}}{r}V_2'(r)\nonumber\\\label{sdpo}
&+&\left(\frac{({\mathbf s_1}{\mathbf r})({\mathbf s_2}{\mathbf
r})}{r^2}-\frac{{\mathbf s_1}{\mathbf s_2}}{3}\right)V_3(r)
+\frac{{\mathbf s_1}{\mathbf s_2}}{3}V_4(r)
\end{eqnarray}
with ${\mathbf s_i}={\mathbf S_i}/M_i, {\mathbf l_i}={\mathbf
L_i}/M_i$
and
\begin{eqnarray}
V_{vd}(r,{\mathbf
p})&=&\frac{1}{8}\left(\frac{1}{M_1^2}+\frac{1}{M_2^2}\right)\nabla^2
\left(V_0(r)+V_a(r)\right)\\
&+&\left\{v_1^i,v_2^j,\left(\delta_{ij}V_b(r)+\left(\frac{\delta_ij}{3}
-\frac{r_ir_j}{r^2}\right)V_c(r)\right)\right\}_W\\
&+&\sum_{k=1}^2\left\{v_k^i,v_k^j,
\left(\delta_{ij}V_d(r)+\left(\frac{\delta_ij}{3}
-\frac{r_ir_j}{r^2}\right)V_e(r)\right)\right\}_W
\end{eqnarray}
with ${\mathbf v_i}={\mathbf p_i}/M_i$,
respectively. $\{\cdot,\cdot,\cdot\}_W$ denotes Weyl ordering of the three
arguments.

The sd and vd potentials $V_1'$, $V_2'$, $V_3$, $V_4$ and $V_a$ --
$V_e$ can be computed from lattice correlation functions
in Euclidean time, $C(T)$. This is done by
measuring expectation values of Wilson loops with
two colour field insertions (ears) within the temporal transporters,
divided by the corresponding loop without ears ($T$ denotes the
ear to ear temporal distance).
One ear excites the gluon field between the two charges while the
second one
returns the field into its ground state. To obtain the  potentials,
an integration over all possible interaction times (temporal
positions of the ears) has to be performed.

The minimal distance of
an ear to an ``end'' of the Wilson loop, occurring
within the integration, $\Delta T$, represents
the time the gluon field has at its disposal to decay
into the ground state after creation and, therefore, governs excited
state contaminations. 
The spatial transporters within the Wilson loops
are smeared\cite{CERN} to suppress such pollutions from
the beginning, allowing us to reduce $\Delta T$.
As an additional technical trick we integrate out temporal links
analytically\cite{Parisi} which results in reduced statistical fluctuations.
Finally, we exploit transfer matrix
techniques to obtain asymptotic results from a finite
integration range (in $T$). Details about this procedure will be
published elsewhere.\cite{inprep}

\section{Matching of the Effective Hamiltonian to QCD}
The sd and vd potentials are computed from amplitudes of
correlation functions
rather than from eigenvalues of the transfer
matrix. This gives rise to renormalizations in respect to
the corresponding continuum potentials. A different way to illustrate the
necessity of renormalization is the fact that
the colour electric (magnetic) ``ears'',
\begin{equation}
gF_{\mu\nu}=\frac{1}{2ia^2}\left(U_{\mu\nu}-U_{\mu\nu}^{\dagger}\right)
+{\cal O}(a^2)\quad,
\end{equation}
which have been inserted into Wilson loops, explicitly depend on the
lattice scale, $a$, and discretization.

However, renormalization is not a pure lattice problem in this case.
By truncating the expansion of the QCD Lagrangian in powers of
$M^{-1}$ at a given order, the ultra-violet behaviour
is changed in respect to the full theory. Therefore, the effective
Lagrangian has to be matched to full QCD at a renormalization scale
$\mu<M$\footnote{On the lattice,
$\mu\simeq\pi/a$.}, giving
rise to renormalization constants $c_i(\mu,M)$, connecting a
QCD potential $V_i(r;M)$ to the corresponding
potential, computed in the framework of
the effective theory, e.g.\ $V_i(r;M)=c_i(\mu,M)V_i(r;\mu,M)$.
This problem, which becomes visible beyond the tree-level,
has been approached systematically for sd potentials in the context of
heavy quark effective theory by Chen {\em et al.}.\cite{Chen}

Up to an additional $\delta^3(r)$-like factor,
that originates from mixing of the dimension 6 spin-spin
interaction term with a four fermion contact term,
a result is obtained which can be rewritten into
Eq.~(\ref{sdpo}) by substituting the naive potentials $V_i$ by
renormalized potentials $V_{i,ren}$ which can be related to each other
by a renormalization constant, $c_3$\footnote{For simplicity we assume
$M=M_1=M_2$.},
\begin{eqnarray}
V_{1,ren}'(r)&=&c_3(\mu,M)V_1'(r,\mu)+\left(c_3(\mu,M)-1
\right)V_0'(r)\quad,\\
V_{2,ren}'(r)&=&c_3(\mu,M)V_2'(r,\mu)\quad,\\
V_{3/4,ren}(r)&=&c_3^2(\mu,M)V_{3/4}(r,\mu)\quad.
\end{eqnarray}

\begin{figure}
\centerline{\mbox{\epsfxsize=12cm\epsfbox{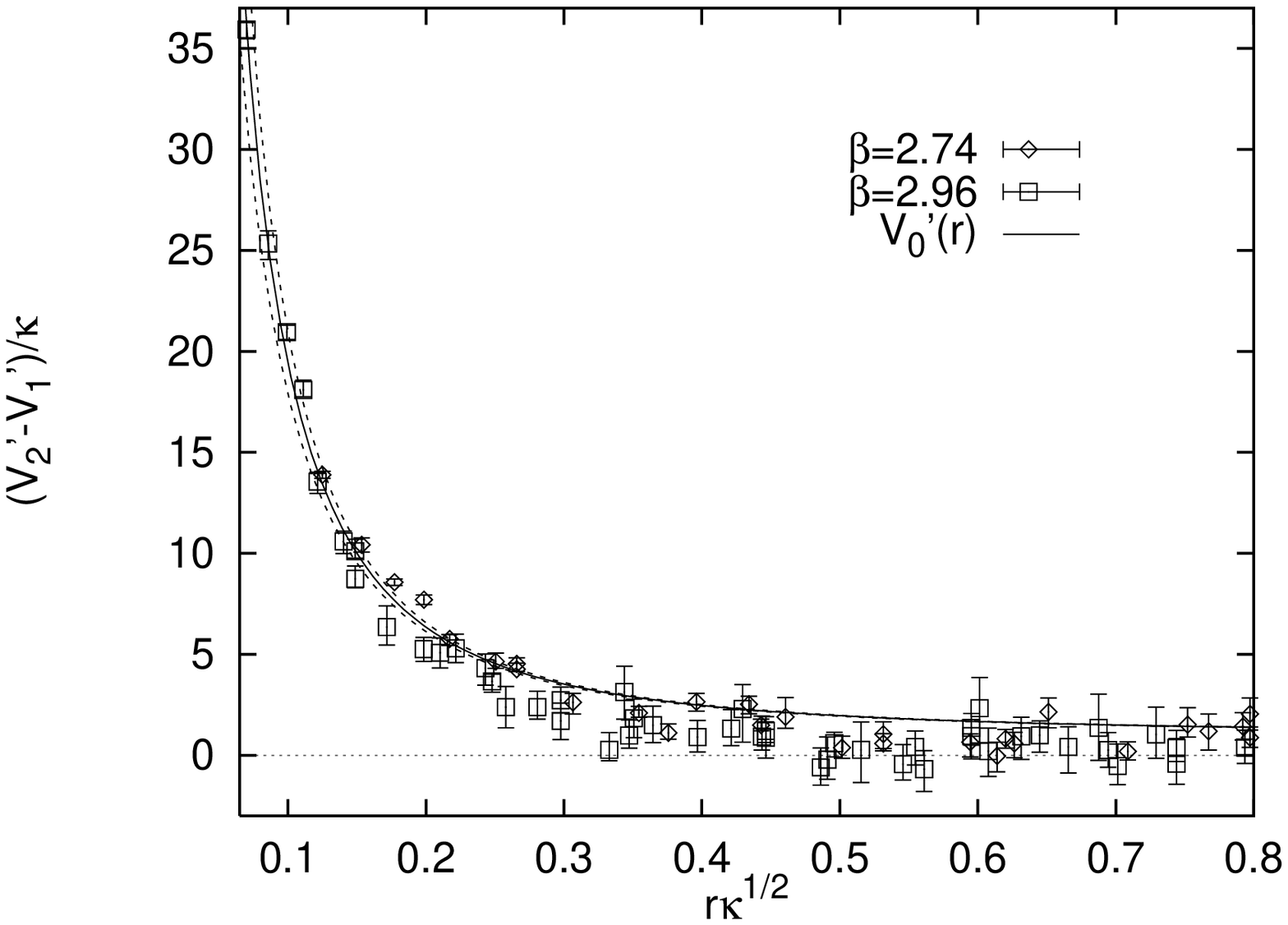}}}
\fcaption{Test of the HM renormalization procedure against the Gromes
relation. The data on $V_2'-V_1'$
correspond to results from two different
lattice spacings. The fit curve refers to the central
force $V_0'$ as computed from smeared Wilson loops.}
\label{fig:gromes}
\end{figure}

The constant $c_3$ is known to one-loop perturbation theory.\cite{Chen}
As $c_3$ is likely to be dominated by higher order perturbative and
nonperturbative uncertainties in the low energy regime of interest, we
apply the (nonperturbative) HM renormalization procedure introduced by
Huntley and Michael.\cite{michael} This procedure has originally been
invented to remove the continuum-lattice renormalization problem
but --- as a by-product --- also
cures the matching problem of the effective heavy quark theory.
The success of this approach can be checked numerically in
two ways, namely (a) by  varying the lattice resolution $a$ and
checking scaling of the results and (b) by testing against the Gromes
relation\cite{Gromes} between spin-orbit potentials and the central
potential
(which does not undergo renormalization)\footnote{
Two additional constraints, relating vd potentials to the central
potential, have been found\cite{VDEP}:
$rV_0'(r)=-2V_c(r)-4V_e(r)$ and
$rV_0'(r)-3V_0(r)=6V_b(r)+12V_d(r)$.},
\begin{equation}
\label{grom}
V_0'(r)=V_{2,ren}'(r)-V_{1,ren}'(r)\quad .
\end{equation}

\begin{figure}
\centerline{\mbox{\epsfxsize=12cm\epsfbox{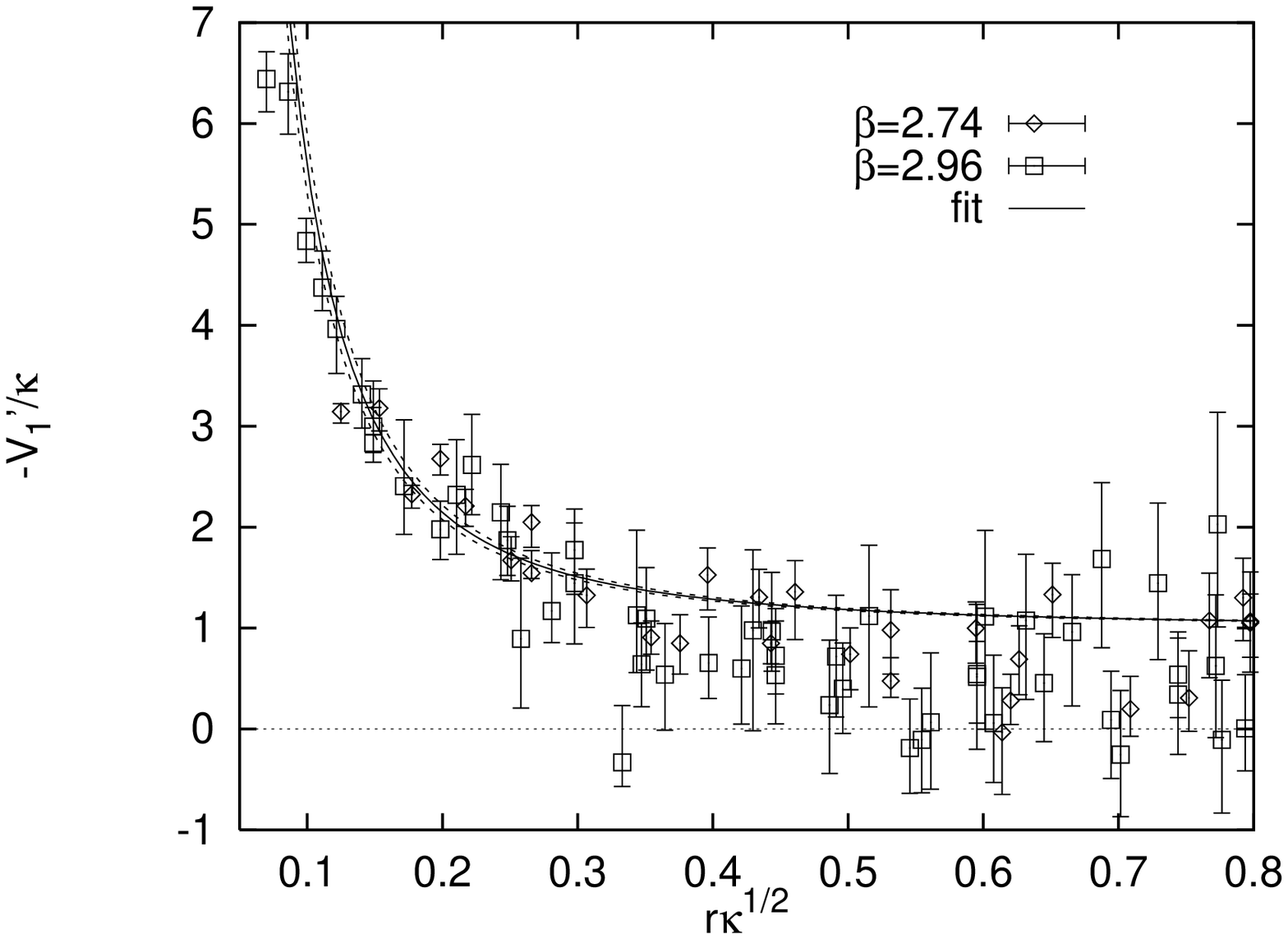}}}
\fcaption{Scaling plot of
the long range spin-orbit potential $V_1'$ (in units of the string
tension, $\kappa$). The curve corresponds to a one-parameter fit of
the form $V_1'(r)=-\kappa-a/r^2$.}
\label{fig:v1}
\end{figure}
\begin{figure}
\centerline{\mbox{\epsfxsize=12cm\epsfbox{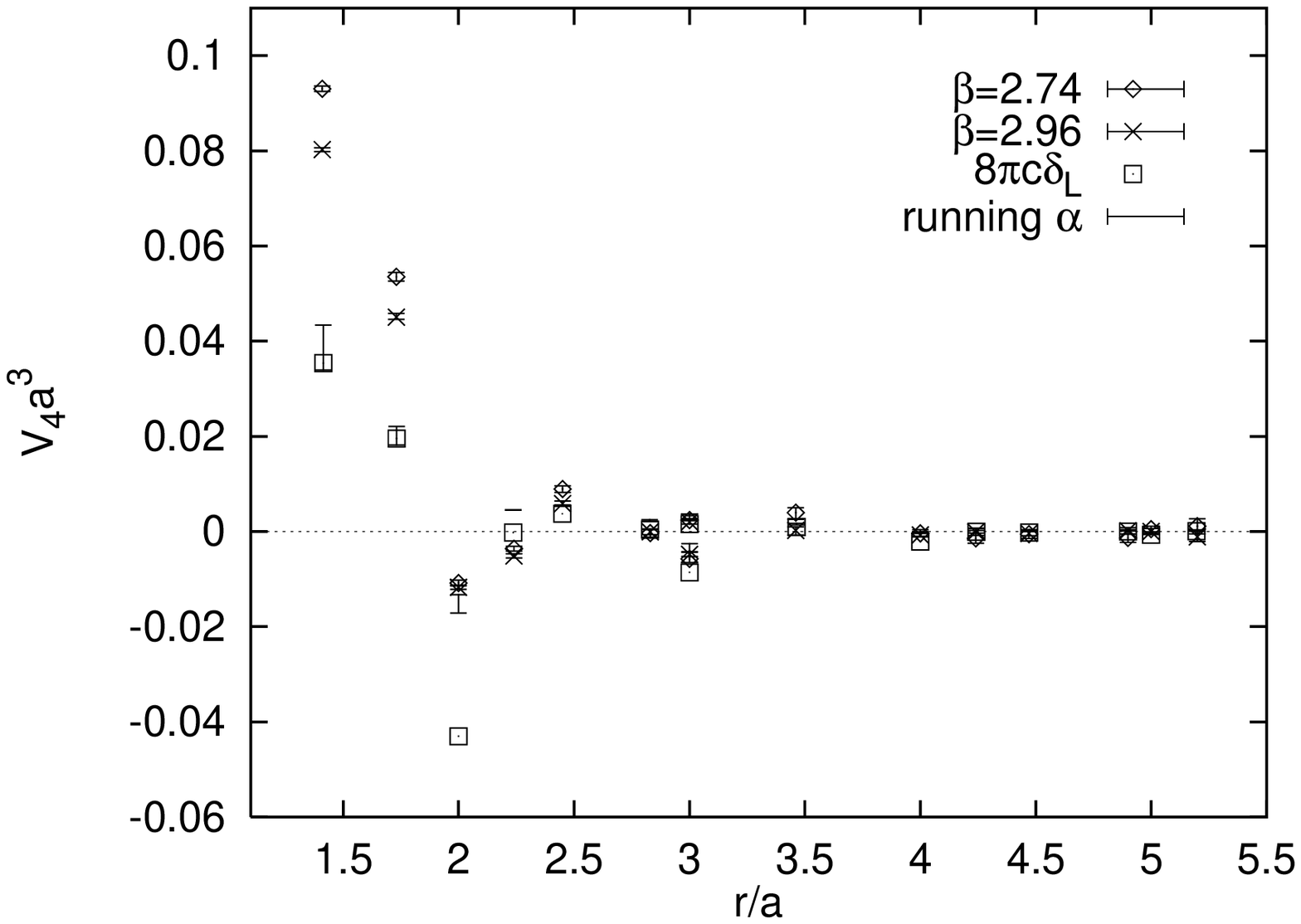}}}
\fcaption{The spin-spin potential $V_4$ in lattice units.}
\label{fig:v4}
\end{figure}
\begin{figure}
\centerline{\mbox{\epsfxsize=12cm\epsfbox{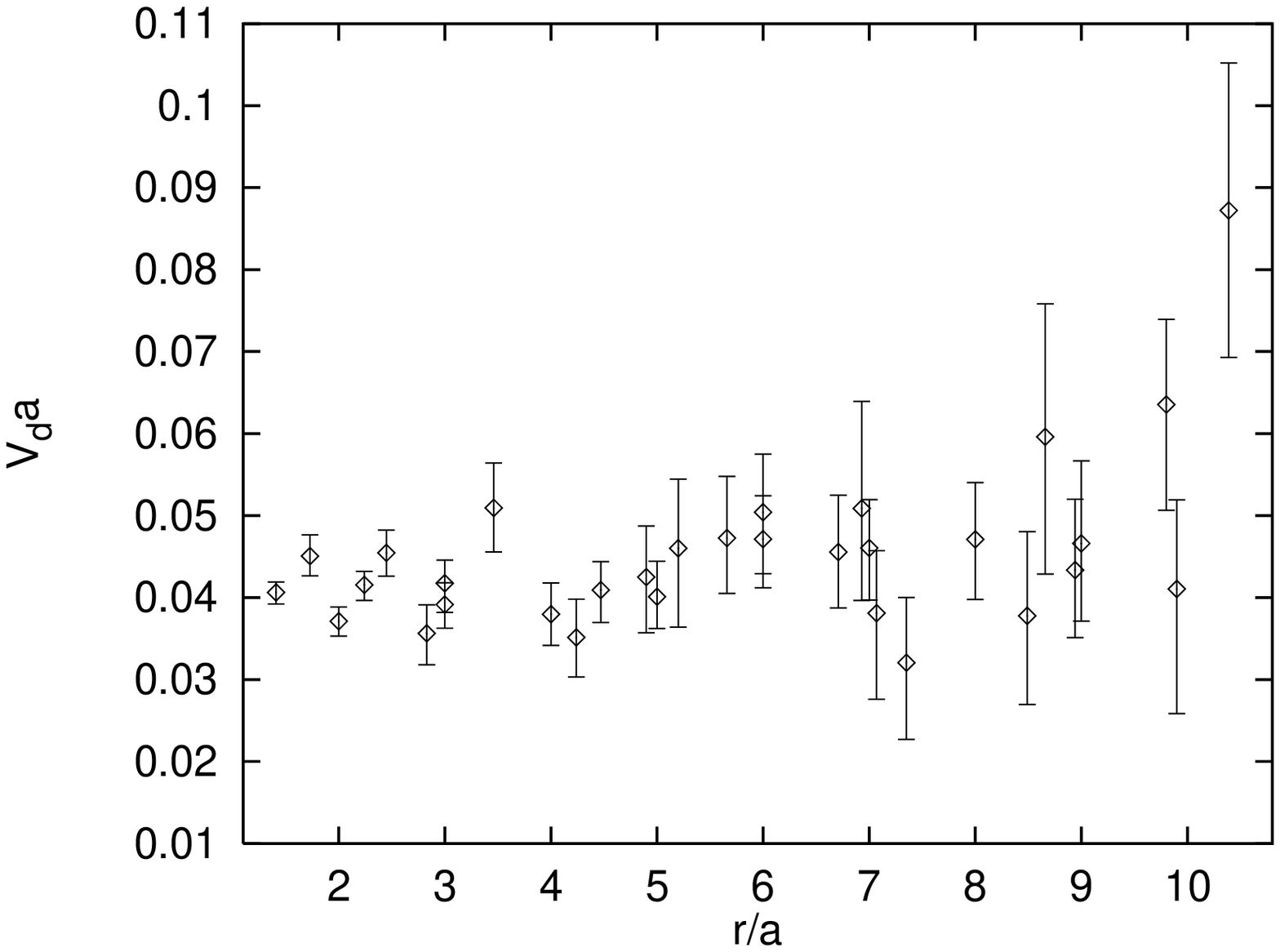}}}
\fcaption{The velocity-dependent potential $V_d$ in lattice units at
$\beta=2.74$.}
\label{fig:vd}
\end{figure}
The present simulations have been performed
on $32^4$ and $48^4$ lattices at $\beta=2.74$ and $\beta=2.96$
which corresponds to lattice spacings $a\approx 0.041$~fm and
$a\approx 0.020$~fm, respectively. 
The above physical scales have been adjusted such that the string tension
comes out to be $\sqrt{\kappa}=440$~MeV.
In Fig.~\ref{fig:gromes} we check our data on $V_2'-V_1'$ (in units of
the string tension, $\kappa$) against the force, obtained from a fit
to the central potential, 
$V_0(r)$. As can be seen, the two data sets scale nicely onto each
other and reproduce the central potential up to lattice artefacts at
small $r$.

\section{Results}
In the continuum, tree-level perturbation theory yields the following
expectations for the spin-orbit and spin-spin potentials,
\begin{equation}
\label{pertu}
V_1'=0\quad,\quad V_2'=\frac{e}{r^2}\quad,\quad V_3=\frac{3e}{r^3}
\quad,\quad V_4(r)=8\pi\delta^3(r)\quad,
\end{equation}
where $e=C_F\alpha$ and $C_F=3/4$ for SU(2). Note, that the first
spin-orbit potential does not contain vector
exchange contributions and, due to the Gromes relation, should be
of the form (assuming Eq.~(\ref{pertu}) to be valid and a Cornell
parametrization of the central potential),
$V_1'=-\kappa$.

The data for $V_1'$  is displayed in Fig.~\ref{fig:v1}.
Our lattice resolution enables us to 
establish an
{\it attractive} short range contribution to $V_1'$ that can be well fitted
to a Coulomb ($1/r^2$) form (in addition to the constant long-range term,
which is in agreement with the string tension, $\kappa$).
This term amounts to
about one quarter of the Coulomb-like contribution to the static potential.

We confirm the second spin-orbit potential $V_2'$ to be definitely
of short range nature. Moreover, up to the first data point, it qualitatively
agrees with tree-level lattice perturbation theory. This agreement can
be made quantitative by allowing for a running coupling parameter.
The same holds true for the spin-spin potential $V_3$.

The remaining spin-spin potential, $V_4$, exhibits oscillatory behaviour as a 
lattice artefact (Fig.~\ref{fig:v4}) and can largely be understood as a 
$\delta$-contribution, according to Eq.~(\ref{pertu}).
The tree-level lattice perturbative expectation $8\pi
c\delta_L^3({\mathbf r}/a)$
is indicated by squares in the figure. The normalization
has been obtained from a
$c/r^2$ fit to the $V_2'$ data points. The error bands without symbols are
obtained by using lattice single gluon exchange with an infra-red
protected two-loop running coupling in momentum space.\cite{inprep} The range
corresponds to different reasonable choices of the QCD $\Lambda$-parameter.
Apart from the dominant $\delta$-like contribution, another very short
ranged contribution seems to exist.

Very recently, we have computed the vd potentials for the first time.
We find reasonable signals for long range forces, as illustrated in
Fig.~\ref{fig:vd}.

\section{Summary and Conclusions}
We have studied central and spin-dependent forces in SU(2) gauge
theory in a high statistics lattice simulation. We find reliable
renormalized potentials with good scaling behaviour. There is clear
evidence for a short range scalar exchange contribution in the first
spin-orbit potential at the level of 20--25 \% of the Coulomb part of the
central potential. The other sd potentials are found to be
short ranged and are well approximated by
perturbation theory.

We are encouraged from the results of a feasibility study of
velocity-dependent potentials.
An extension of the present investigations to the case of
interest, SU(3) gauge theory, is in progress.

\section{Acknowledgements}
We thank DFG for supporting the
Wuppertal CM-2 and CM-5 projects (grants Schi 257/1-4 and Schi
257/3-2). We are grateful for computing
time on the CM-5 at GMD in Birlinghoven and appreciate support 
by EU (grant CHRX-CT92-0051).

\end{document}